\documentclass[journal]{IEEEtran}
\usepackage{cite}
\usepackage{graphics} 
\usepackage{csquotes}
\usepackage{amsmath}
\usepackage{amssymb,amsfonts}
\usepackage{algorithmic}
\usepackage{graphicx}
\usepackage{textcomp}
\usepackage{indentfirst}
\usepackage{graphicx}
\usepackage{subfigure}
\usepackage{csquotes}
\usepackage{multirow}
\usepackage{tabularx}
\usepackage{float}
\usepackage{gensymb}
\usepackage{commath}
\usepackage{scalerel,stackengine}
\usepackage[normalem]{ulem}
\usepackage{color}
\usepackage{dblfloatfix}
\usepackage[table,xcdraw]{xcolor}
\usepackage{authblk}
\usepackage{siunitx}
\usepackage{soul}
\usepackage{lipsum}
\usepackage{algorithmic}
\usepackage[linesnumbered,ruled,vlined]{algorithm2e}

\DeclareMathOperator{\sgn}{sgn}

\newcommand{\Vthset}{\ensuremath{V_{\mathrm{th}}^{\mathrm{(set)}}}}

\newcommand{\Vthreset}{\ensuremath{V_{\mathrm{th}}^{\mathrm{(reset)}}}}

\SetCommentSty{mycommfont}

\begin{document}
\title{Memristor Hardware-Friendly Reinforcement Learning}
\author{Nan Wu,
        Adrien Vincent,
        Dmitri Strukov,
        and Yuan Xie}

\maketitle

\begin{abstract}
Recently, significant progress has been made in solving sophisticated problems among various domains by using reinforcement learning (RL), which allows machines or agents to learn from interactions with environments rather than explicit supervision. 
As the end of Moore's law seems to be imminent, emerging technologies that enable high performance neuromorphic hardware systems are attracting increasing attention.
Namely, neuromorphic architectures that leverage memristors, the programmable and nonvolatile two-terminal devices, as synaptic weights in hardware neural networks, are candidates of choice to realize such highly energy-efficient and complex nervous systems. 
However, one of the challenges for memristive hardware with integrated learning capabilities is prohibitively large number of write cycles that might be required during learning process, and this situation is even exacerbated under RL situations.

In this work we propose a memristive neuromorphic hardware implementation for the actor-critic algorithm in RL. 
By introducing a two-fold training procedure (i.e., \textit{ex-situ} pre-training and \textit{in-situ} re-training) and several training techniques, the number of weight updates can be significantly reduced and thus it will be suitable for efficient in-situ learning implementations. 
Two different scenarios are considered: (1) with complete environmental information, the re-training can start with a rather good situation and adjust functionality based on specific artifacts; 
(2) with limited environmental information, we propose an off-policy method with experience replay and importance sampling in pre-training, and a synchronous parallel architecture in re-training taking advantages of both parallelism and the increase of sample efficiency. 
As a case study, we consider the task of balancing an inverted pendulum, a classical problem in both RL and control theory. 
We believe that this study shows the promise of using memristor-based hardware neural networks for handling complex tasks through in-situ reinforcement learning.
\end{abstract}


\section{Introduction}
\IEEEPARstart{M}{any} ideas in artificial intelligence are based on supervised learning, in which machines or agents are trained to mimic decisions made by humans.
However, under certain circumstances the training data sets are expensive or considerably difficult to obtain~\cite{lecun2015deep}.
For example, in control problems or many sequential decision-making processes, it is nearly impossible to figure out the explicit supervision.
Reinforcement learning (RL) allows, by contrast, machines or agents to simply learn from interacting with environments, through rewards or punishments based on their behaviors.
Such an approach based on explicit environmental information yields a powerful machine-learning framework. For example, Mnih \textit{et al.} demonstrated a deep Q-network agent that bridges the gap between high-dimension sensory inputs and actions, capable of learning challenging tasks, and competing or even surpassing professional human players~\cite{mnih2015human}.
Silver \textit{et al.} introduced AlphaGo Zero, with neural networks trained by reinforcement learning from self-play, exceeding the human's capabilities~\cite{silver2017mastering}.
Other investigations in real-life control problems, from robotic arm manipulation to self-driving problems~\cite{james20163d,gu2017deep,levine2018learning,sallab2017deep}, highlight the versatility of reinforcement learning.
    
Recently, both academia and industry have shown an increasing interest for neuromorphic \emph{hardware} systems, which could eventually be faster and more energy-efficient than their software counterparts that are typically implemented on conventional microprocessors or graphics processing units. 
However, neuromorphic systems based on purely conventional silicon technologies may only provide limited long-term potentials and capabilities, as Moore's law seems to be coming to an end and digital CMOS circuits already operate close to their physical limits.
At the same time, various types of emerging nonvolatile memories, which can lead to breakthroughs in both memory and computing architectures, are being investigated for their use in fast and energy-efficient neuromorphic networks.
One of the most prospective candidates is metal-oxide memristors ~\cite{yang2013memristive} (also referred as resistive switching memories or ReRAMs), which are programmable analog nonvolatile memory devices that can be scaled down below \SI{10}{\nm} without sacrificing their performance ~\cite{hu2014memristor,bayat2018implementation,prezioso2015training}.  
For memristive hardware with integrated learning capabilities, one of the major challenges is prohibitively large number of write cycles that might be required during learning process, and this situation is even exacerbated under RL situations.

In this work, we propose a neuromorphic hardware implementation for reinforcement learning based on memristive circuits. We specifically focus on the inverted pendulum problem, a well-known and commonly used benchmark for reinforcement learning algorithms~\cite{wu2017scalable,lillicrap2015continuous,schulman2015trust,schulman2017proximal}. Indeed, the inverted pendulum problem is often considered as an instance of the inherently unstable dynamic systems existing in various control problems, where many researchers are interested to handle with emerging technologies.
For example, Gale \textit{et al.} designed a memristor-model based hybrid robot control system ~\cite{gale2014design}; carbon nanotube devices were used to implement a robot controller\cite{mohid2015evolving} and a handshaker\cite{shulaker2013sacha}; Dong \textit{et al.} proposed to construct a quantum robot~\cite{dong2006quantum}. 

The specific contributions of our work are summarized as follows:
\begin{itemize}
    \item We introduce a two-fold reinforcement learning procedure: \textit{ex-situ} pre-training and \textit{in-situ} re-training, which can significantly reduce the writing cycles required during the \textit{in-situ} learning process.
    \item Based on the two-fold learning procedure, we propose a memristive neuromorphic hardware-friendly implementation that is suitable for \textit{in-situ} reinforcement learning.
    \item Two different scenarios are considered based on the difficulty to obtain environmental information in pre-training:
    \begin{itemize}
        \item If environmental information is complete in pre-training, the re-training can start with a rather good pre-training and further adjust functionality based on specific artifacts (see Section III.A).
        \item If environmental information is limited in pre-training, we propose an off-policy method with experience replay and importance sampling in pre-training, and a synchronous parallel architecture in re-training (see Section III.B).
    \end{itemize}
    \item We develop novel training approaches with significantly fewer number of weight updates to address the switching endurance problem of memristors. We also modify the temporal difference learning algorithm to compensate for the loss of functional performance due to nonlinear switching dynamics and device to device variations typical for metal oxide memristors.
\end{itemize}

We introduce the inverted pendulum problem and the actor-critic based reinforcement learning in Section II. The proposed training methods and the simulation framework are discussed in Section III. In section IV, hardware implementation and architecture design are detailed. Simulation results are discussed in Section V, and we summarize our work in Section VI.

\section{Preliminary}

\subsection{Inverted Pendulum Problem and Actor-Critic Algorithm}
\begin{figure}[btp]
\centering
	\includegraphics[width=2.3in]{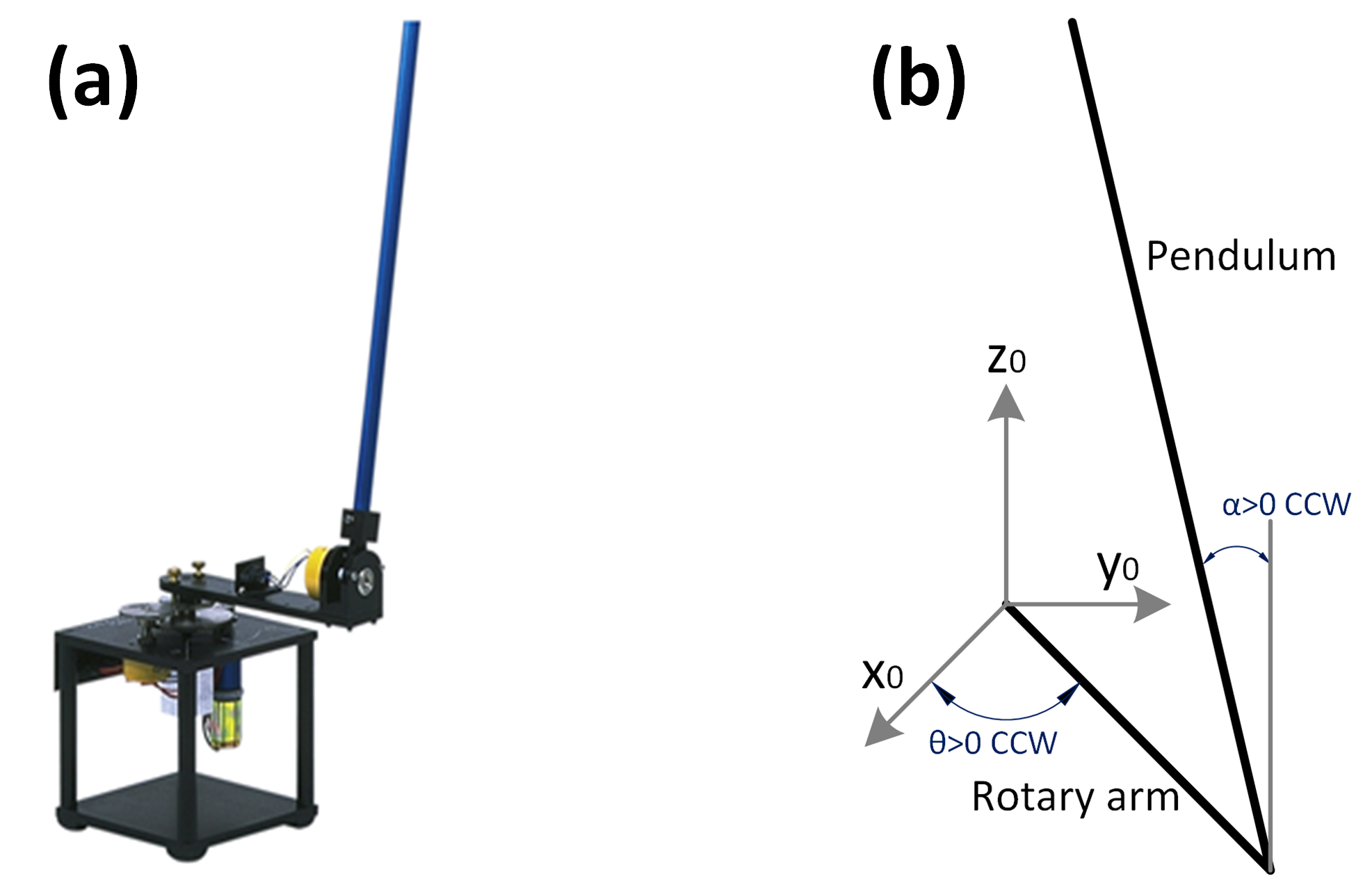}
    \caption{(a) Rotary inverted pendulum and (b) its modeling scheme.}
    \label{fig:pendulum}
\end{figure}

We study a rotary inverted pendulum problem using parameters of a Quanser SRV02 system \cite{quanser2011, quanser2009} as a case study. In such a system, the pendulum is connected to the end of the rotary arm, which is electrically actuated (Fig.~\ref{fig:pendulum}).  The general goal is to keep the pendulum in the defined upright region via sequential application of clockwise (CW) or counter-clockwise (CCW) pushes to the rotary arm at discrete time steps. In the setup, the rotary arm is pushed by applying fixed magnitude voltage pulses to the arm's actuator. Note that no push is not allowed.

The inverted pendulum problem is modeled using Markov decision process. At each time step $t$, the agent receives a state $s_t$ of the rotatory inverted pendulum, which is characterized by four parameters: the rotary arm angle $\theta$, its angular velocity $\dot{\theta}$, the pendulum angle $\alpha$, and its angular velocity $\dot{\alpha}$. Note that angle $\theta$ increases when rotating CCW, while angle $\alpha$ is zero when pendulum is perfectly upright and increases when arm is rotated CCW. After receiving the state $s_t$, the agent selects an action $a_t$ from the binary action space (i.e., $a_t=1$ for a CCW push or $a_t=0$ for a CW push) according to its policy $\pi$, where $\pi$ is a mapping from states $s_t$ to actions $a_t$. In return, the agent receives the next state $s_{t+1}$ and a scalar reward $r_t$. Regarding the task to balance the pendulum, the reward $r_t$ is defined as

\begin{equation}\label{eq:r}
    r_t=\left\{
    \begin{array}{c l}
         0, & \mathrm{if~}\abs{\alpha_t} < 10\degree \\
         -1, & \mathrm{otherwise.}
    \end{array}\right. 
\end{equation}

The return $R_t={\sum\limits_{k=0}^\infty \gamma^kr_{t+k+1}}$ is the total accumulated rewards at the time step $t$ with a \textit{discount rate} (DR) $\gamma\in[0,1)$. The value of state $s$ under policy $\pi$ is defined as $V_{\pi}(s)=\mathbb{E_\pi}\lbrack R_t|s_t=s\rbrack$, which is the expected return for following policy $\pi$ from state $s$. The goal of the agent is to maximize $V_{\pi}(s)$ for each state $s$.

The inverted pendulum problem is solved by the actor-critic algorithm with one-step temporal difference method \cite{barto1983neuronlike,kimura1998analysis,sutton1984temporal,anderson1987strategy, sutton1998reinforcement}, where the policy $\pi$ is the actor and the state value function $V_{\pi}(s)$ is the critic. Let $V_{\pi}(s;\theta_v)$ and $\pi(a|s;\theta_p)$ be an approximate state value function with parameters $\theta_v$ and an approximate policy with parameters $\theta_p$, respectively. The one-step temporal difference is defined as

\begin{equation}\label{eq:tde}
    \delta_t=r_{t}+\gamma {V_\pi}(s_{t+1};\theta_v)-{V_\pi}(s_{t};\theta_v). 
\end{equation}

Then the value parameters $\theta_v$ are updated by semi-gradient descent on the $l2-norm$ of $\delta_t$, which is in the direction of

\begin{equation}\label{eq:update_v}
    \delta_t\nabla_{\theta_v}V_{\pi}(s_t;\theta_v).
\end{equation}
And the policy parameters $\theta_p$ are updated by gradient ascent on $\delta_t$, which is in the direction of the policy gradient:

\begin{equation}\label{eq:update_p}
    \delta_t \nabla_{\theta_p}\log \pi(a_t|s_t;\theta_p)
\end{equation}

\subsection{Memristive Devices}
\begin{figure}[btp]
\centering
    \includegraphics[width=3.4in]{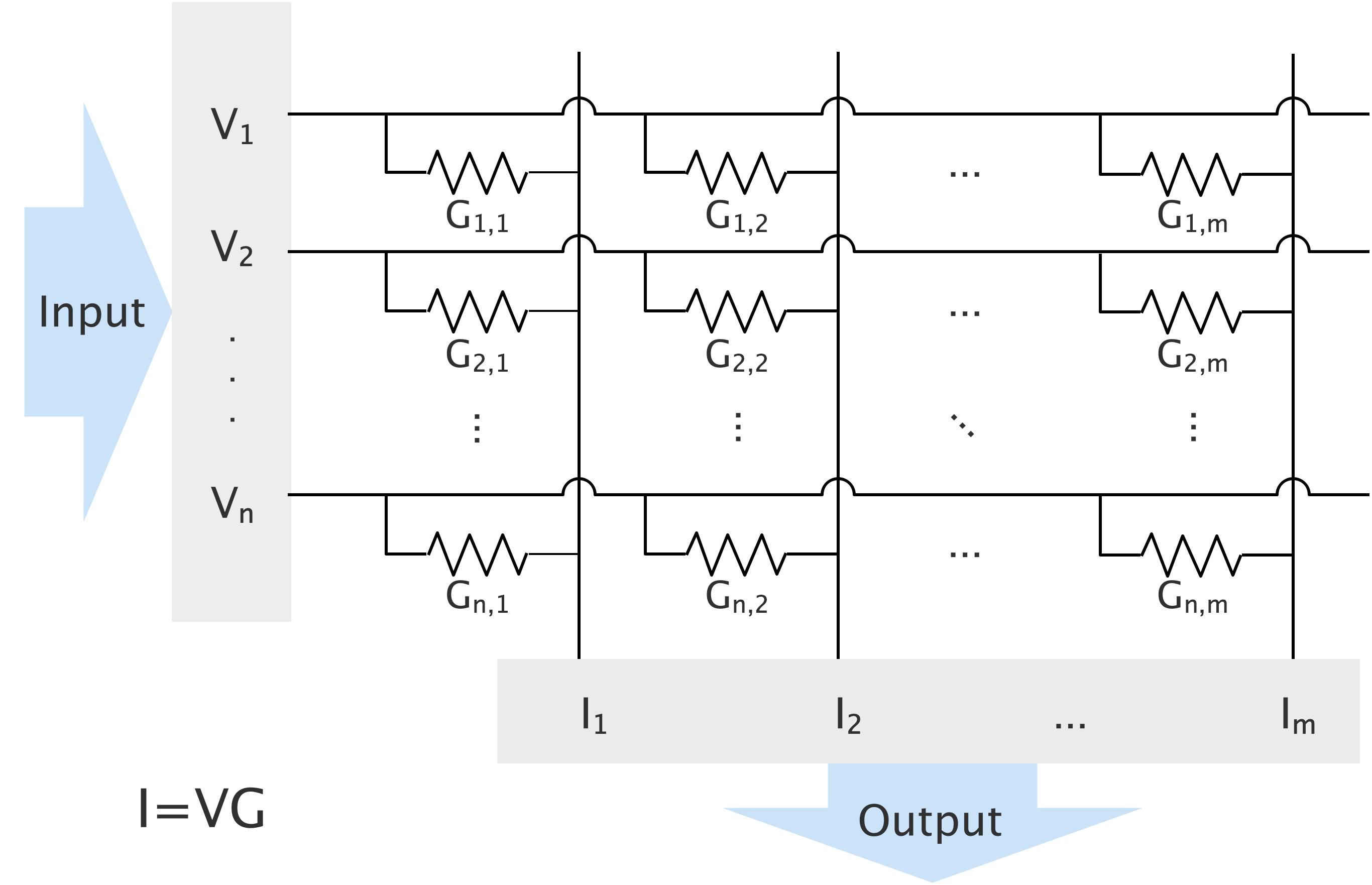}
    \caption{Vector-matrix multiplication with the memristor crossbar.}
    \label{fig:vmm}
\end{figure}

With long retention, high integration density, low power consumption, analog switching and high non-linearity of current-voltage curve, memristive devices are considered as very promising candidates for the implementation of artificial synapses~\cite{strukov2012resistive}.
Additionally, since memristors can integrate computation and memory in the same physical place, data movements between computing and memory elements can be significantly reduces. With these characteristics, memristor-based crossbars are very efficient at computing analog vector-matrix multiplications in the locations where the matrices are stored, highlighting its potentials in neuromorphic computing. 
As shown in Fig.~\ref{fig:vmm}, there is a memristor cell in each intersection of the crossbar. An input vector $V$ is applied to the rows and multiplied by the conductance matrix of memristors. The resulted currents are summed up by Kirchhoff's law in each column, deriving the output current vector $I=VG$.

We consider the device model for $\mathrm{Al}_2\mathrm{O}_3/\mathrm{TiO}_{2-x}$ memristors from \cite{prezioso2015training}. According to that model, the change in the conductance is determined by $slope$, $\Vthset$ and $\Vthreset$ in which $slope$ is a function of the pulse amplitude and duration, $\Vthset$ and $\Vthreset$ are the effective set and reset switching threshold voltages, respectively. We assume three cases for simulations of memristors: (i) ideal devices with no variations and $\Vthset=\Vthreset=$ \SI{2.25}{\V}; (ii) non-ideal devices with $\Vthset$ and $\Vthreset$ randomly chosen within  $\pm$\SI{30}{\percent} of the ideal device's thresholds; and iii) non-ideal devices with both set and reset thresholds uniformly distributed within $[\SI{1}{\V}, \SI{5.5}{\V}]$ range, as discussed in Ref.~\cite{prezioso2015training}.
\section{Method}
\begin{figure}[tbp]
\centering
    \includegraphics[width=3.4in]{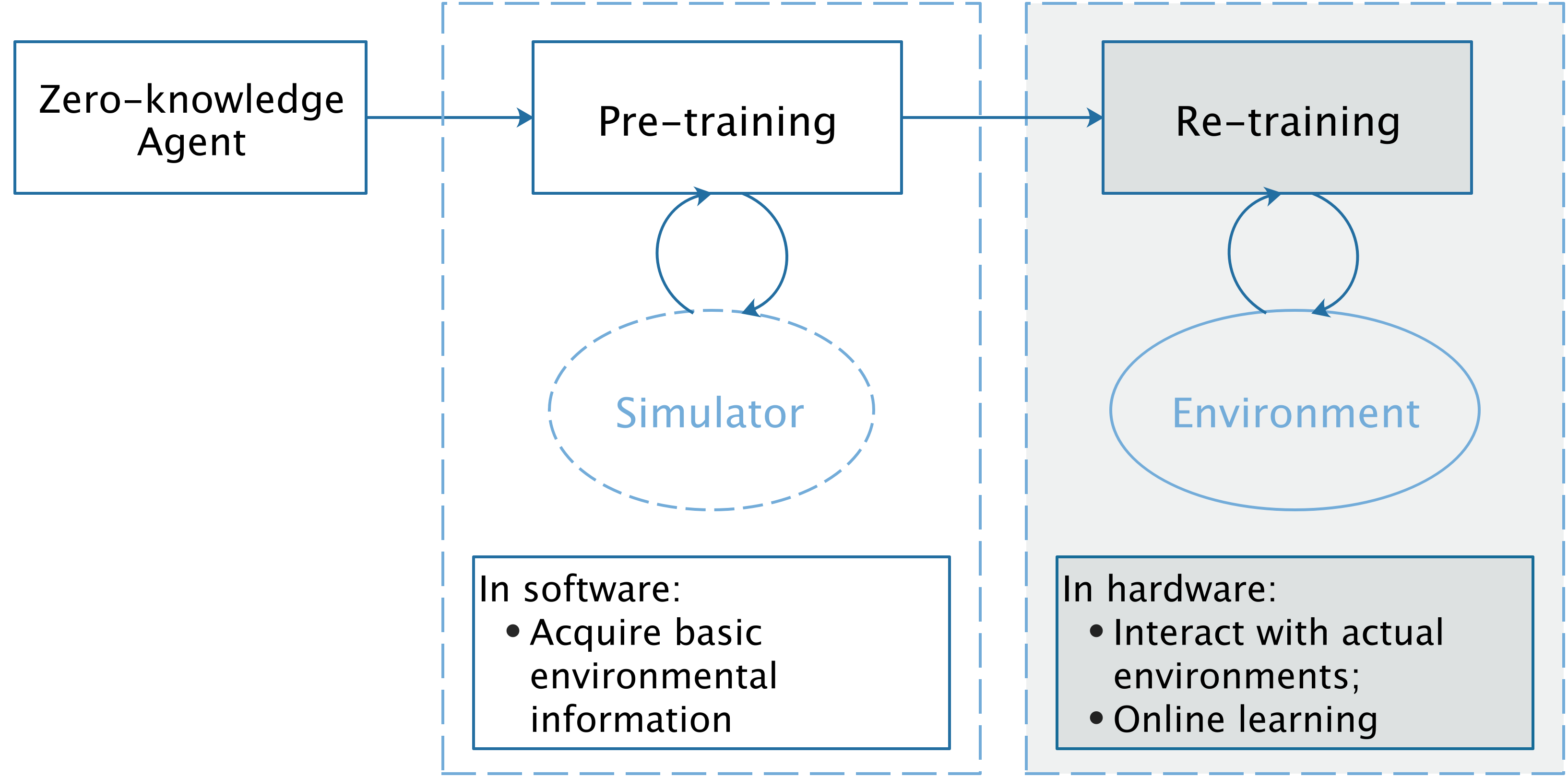}
    \caption{Two-fold training procedure: pre-training \textit{ex-situ} and re-training \textit{in-situ}.}
    \label{fig:twofold}
\end{figure}

There are many many advantages of \textit{in-situ} reinforcement learning with memristive neuromorphic hardware, e.g., higher energy efficiency, speedup, mitigation of process variation and better adaptivity of environments.
However, there are several challenges needed to be considered. 
First of all, in most RL situations huge amount of training data are required by the agent to converge, which may wear our the endurance of memristive hardware. 
Second, the safety issue is also a concern, as there is no supervision during the \textit{in-situ} training, the agent may do something harmful to itself or to the environment. 
Based on these considerations, we propose a two-fold training procedure: \textbf{\textit{ex-situ} pre-training and \textit{in-situ} re-training}. 
As shown in Fig.~\ref{fig:twofold}, in the pre-training phase the agent is trained \textit{ex-situ} to acquire basic environmental information; 
and in the re-training phase the agent is trained \textit{in-situ} by direct interactions with actual environments starting from pre-trained weights. 
With such a procedure, fewer samples will be required in \textit{in-situ} training, which looses the requirements of hardware endurance during training; 
agents will take safer behaviors as they already have some knowledge about the environment; 
and this procedure can still benefit from \textit{in-situ} training.

Two scenarios are considered with respect to the difficulty to obtain environmental information: 
if environmental information is easy to obtain in pre-training, the re-training can start with a rather good pre-training and further adjust functionality based on specific artifacts (detailed in Section III.A); 
if environmental information is expensive or difficult to obtain in pre-training, we propose an off-policy method with experience replay and importance sampling in pre-training, and a synchronous parallel architecture in re-training (detailed in Section III.B).

\subsection{Complete Environmental Information in Pre-training}

In this scenario, we assume that the simulator can get complete environmental information in pre-training. 
Thus, in the pre-training phase, the agent is trained \textit{ex-situ} for balancing the averaged pendulum with standard length and mass (0.127 kg mass and 0.3365 m length) starting from random initial weights; 
in the next, the re-training phase, the agent is trained \textit{in-situ} for balancing pendulums with variations, starting from initial pre-trained weights. 
In fact, this corresponds to a situation that is likely to happen in real life when agents would be pre-trained to solve some average-configuration task and then will have to adjust their functionality based on the specific artifacts of a given task.

\subsubsection{Training Approach}
\begin{figure}[tbp]
\centering
    \includegraphics[width=3in]{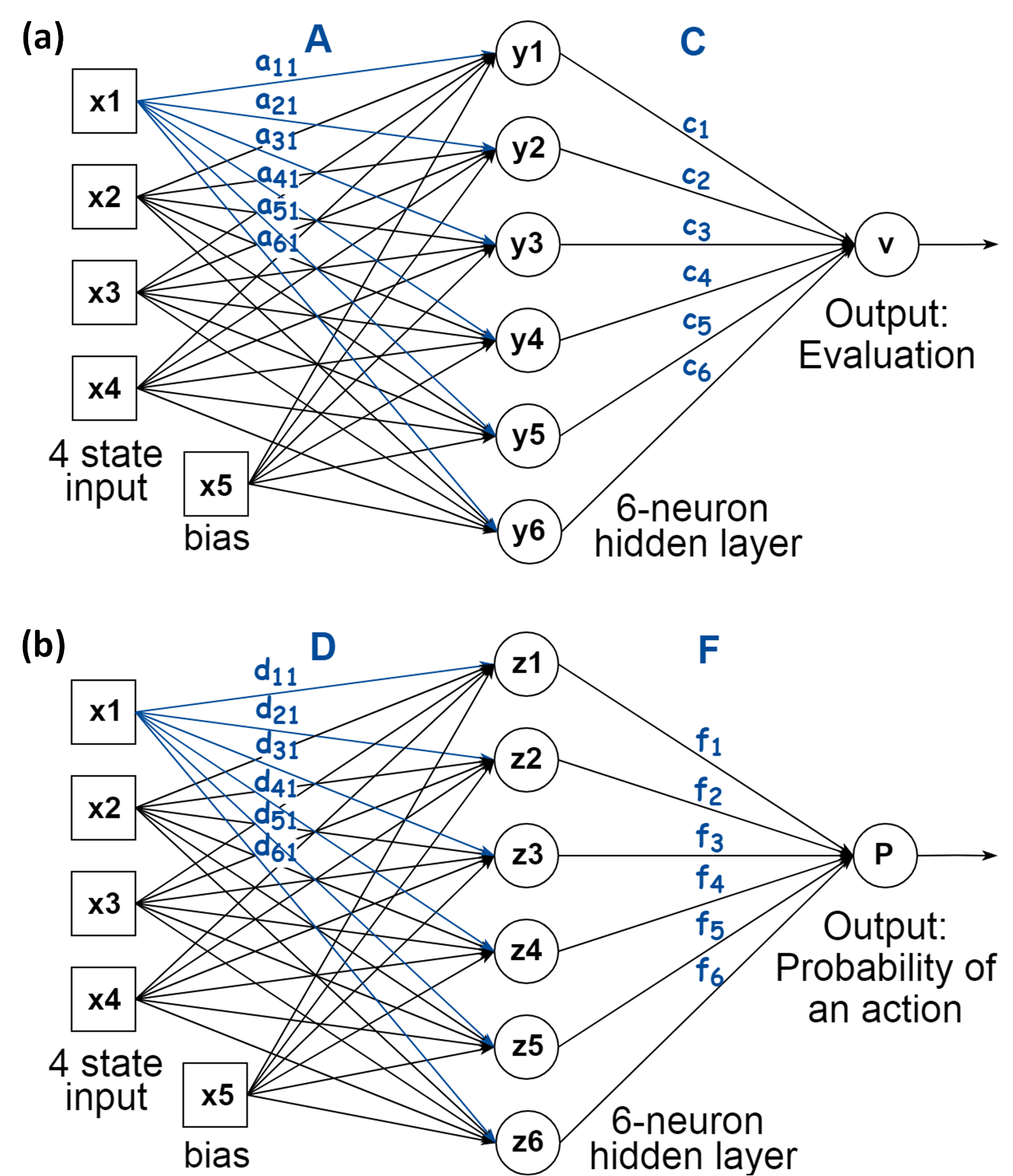}
    \caption{(a) Evaluation and (b) action neural networks.}
    \label{fig:nn}
\end{figure}

To implement the actor-critic algorithm, we employ the action (`actor') and evaluation (`critic') networks. 
The action network implements a stochastic action policy $\pi(a|s;\theta_p)$, and the evaluation network approximates the state-value function with $V_{\pi}(s;\theta_v)$.
These two neural networks have the same structure with 5 inputs, a 6-neuron hidden layer, and 1 output (Fig.~\ref{fig:nn}).
Their five inputs represent the four state parameters, normalized to their maximum values, and a bias term. 
Based on Eq.(\ref{eq:update_v}) and Eq.(\ref{eq:update_p}), the network weights, using notations from Fig.~\ref{fig:nn}, are modified as 
{
\begin{equation}
    \begin{split}
    c_i(t+1)    & = c_i(t)    + \beta\delta_t y_i(t), \\
    a_{ij}(t+1) & = a_{ij}(t) + \beta_h\delta_t y_i(t)(1-y_i(t))\sgn(c_i(t))x_j(t), \\
    f_i(t+1)    & = f_i(t)    + \rho\delta_t(q(t)-p(t))z_i(t), \\
    d_{ij}(t+1) & = d_{ij}(t) \\
                &+\rho_h\delta_t  (q(t)-p(t))z_i(t)(1-z_i(t))\sgn(f_i(t))x_j(t),
    \end{split}
\label{eq:modify}
\end{equation}
}where $\beta$, $\beta_h$, $\rho$ and $\rho_h$ are layer-specific learning rates, and we let $q(t)=a_t$ and $p(t)=\pi(a_t|s_t;\theta_p)$.

To evaluate our idea and study the potential performance for its hardware implementation we consider three main training approaches: (i) \emph{Baseline} that does not involve any pre-training, (ii) \emph{Exact}, and (iii) \emph{Manhattan} in which we use pre-training. 
In the first two approaches, weights are updated exactly by the prescribed amounts at each time step according to the Eq.~(\ref{eq:modify}), while in \emph{Manhattan}, which is a representative of the hardware implementation, only the sign of the update is considered. 

Also, for each of these approaches, we consider a slight modification for the training algorithm, denoted with the additional label `PQ', in which weights are updated only at time steps when $\abs{p-q}$ is larger than 0.9 for \emph{Baseline} and \emph{Exact}, and 0.95 for \emph{Manhattan}. 
Note that in the re-training case, $\delta_t$ is typically small, thus in PQ training, the number of updates is reduced by only performing significant weight updates.   

For all cases, $\rho$=0.25, $\rho_h$=0.2, $\beta$=0.2, $\beta_h$=0.1. $\gamma$ is 0.85, 0.9,  0.75 for \emph{Baseline}, \emph{Exact}, and  \emph{Manhattan}, repsectively.  
The slope of the sigmoid function of the output neuron in action network is eight times of that of hidden neurons, to make weights in both layers comparable.

Finally, for the \emph{Manhattan} approach, we also consider adjusting the discount rate $\gamma$ during training for each agent.
This is performed by changing $\gamma$ by 0.02 each 50 trials (see definition of a trial in the next subsection), when the successful trial rate is less than 35\%, in the direction determined based on the observed performance.

\subsubsection{Simulation Framework}

For our simulations, we consider 25 pendulum configurations, each with a different combination of pendulum's mass and length, which are selected from  \{\SI{-10}{\percent}, \SI{-5}{\percent}, \SI{0}{\percent}, \SI{5}{\percent}, \SI{10}{\percent}\} of standard pendulum values.
To collect statistics, for each pendulum, we consider 100 different sets of initial weights for pre-training, and similarly, 100 different sets of pre-trained weights, all obtained after pre-training on the standard pendulum configuration. This results in 2500 total simulated scenarios, which are referred as `agents', for each training approach.

Furthermore, we use randomly generated initial states of the pendulum in the upright region, with the total of 7000, 2000, and 500 different initial states for pre-training, re-training, and testing, respectively. The training or testing of each agent is performed in a sequence of trials. In each trial, the pendulum is set to an initial position, which is randomly chosen from the corresponding (i.e. pre-training,  re-training, or testing) set. During each trial, the pendulum is balanced and, in the case of pre-training and re-training, the weights are updated according to the chosen training approach. The trial is ended when either the pendulum falling down (a failure trial), or successfully kept upright for 5000 time steps (a successful trial). In the first trial, the initial weights are selected according to the specific agent, while the initial weights for the remaining trials correspond to those at the end of the immediately preceding trials. A sequence of trials is continued until there are $C$ total successful trials. A larger $C$ effectively means a longer training time. The total number of trials never exceeds 7000 (2000) for pre- (re-) training.

Besides the total number of updates per weight, which related to the endurance of memristive hardware, other metrics of interest are time steps to failure (\textit{t2f}) and efficiency of weight updates. The former measures the length of time, in time steps, for keeping the pendulum upright. The latter is defined as an improvement in \textit{t2f} with respect to that of pre-trained network, divided by the number of performed updates per weight for a particular training approach.

\begin{figure}[]
\centering
    \includegraphics[width=3.3in]{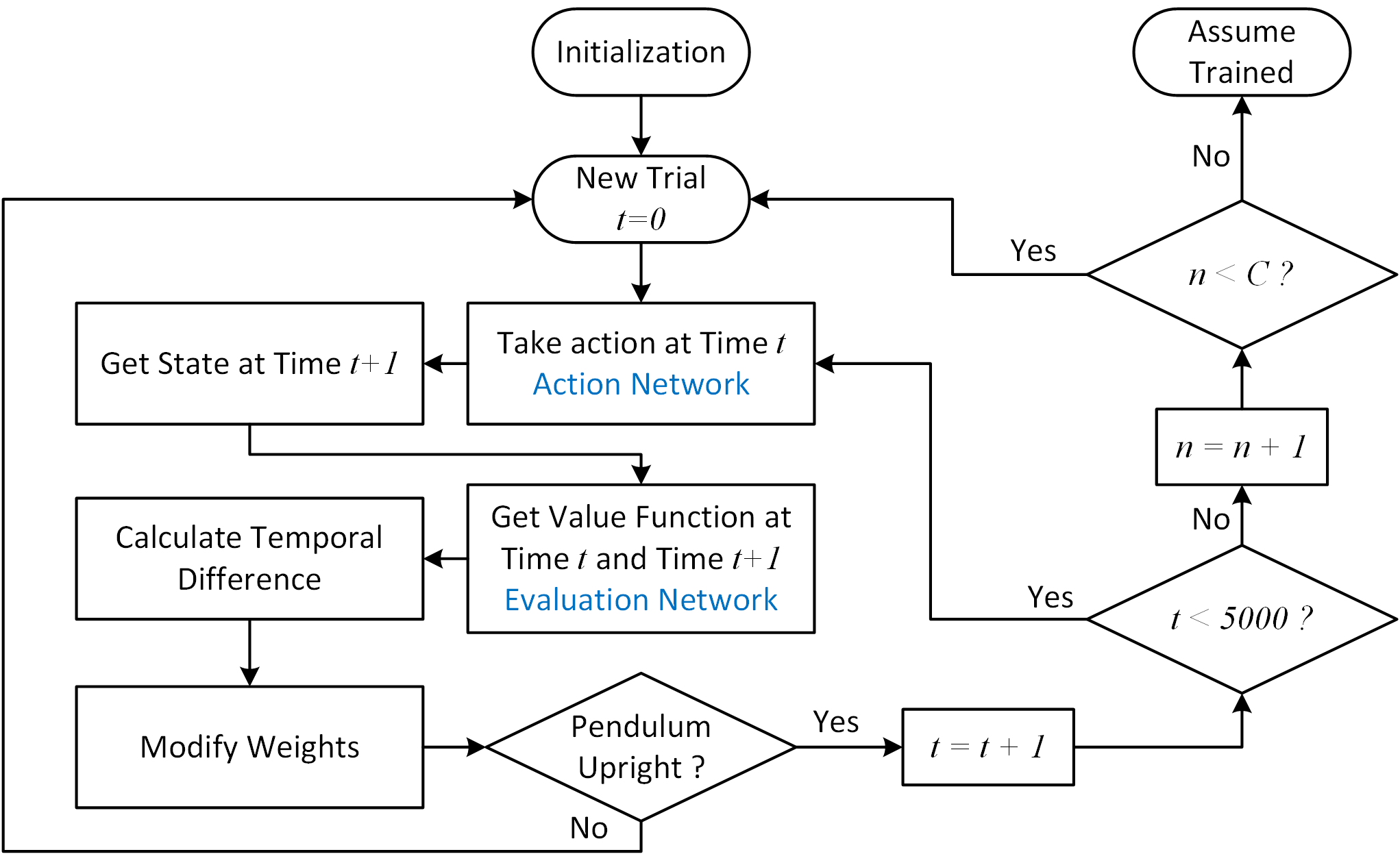}
    \caption{Simulation flowchart for the complete environmental information scenario. Here $t$ is a \emph{time step} that is 0.02s, $n$ is a current number of successful trials. $C$ is the selected criterion to stop training.}
    \label{fig:flowchart}
\end{figure}

Simulation workflow is summarized in Fig. ~\ref{fig:flowchart}.

\subsection{Limited Environmental Information in Pre-training}
In those very complicated environments, it is expensive or nearly impossible to obtain complete environmental information. 
In this scenario, we assume that only limited environmental information is available in the pre-training phase, i.e., discrete samples of environments in tuples $(s_t, a_t, r_t, s_{t+1})$ are available in the pre-training.
Then in the re-training phase, the agent is trained \textit{in-situ} with the standard pendulum from pre-trained weights.

\subsubsection{Off-policy Pre-training with Experience Replay and Importance Sampling}
\begin{figure}[tbp]
\centering
    \includegraphics[width=3in]{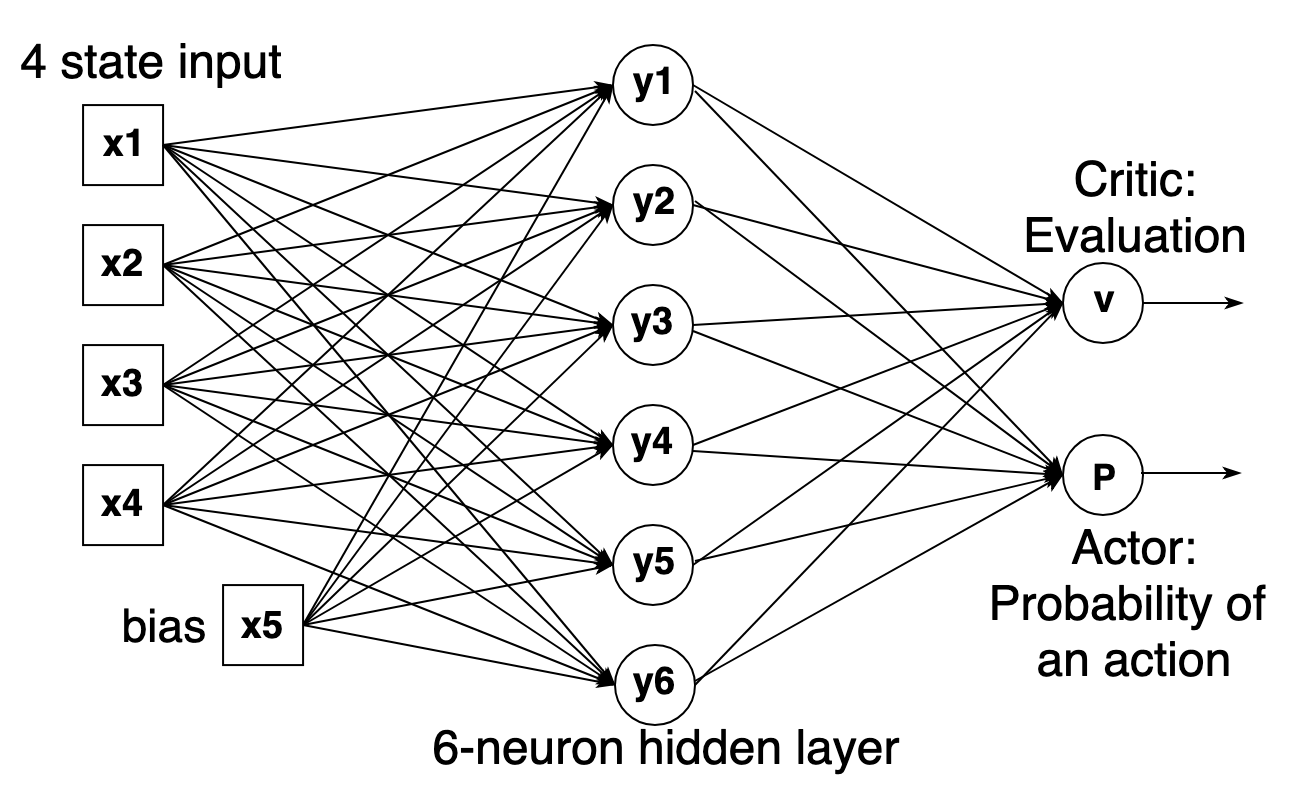}
    \caption{Actor and critic share one neural network.}
    \label{fig:1nn}
\end{figure}

We make the actor and the critic share one neural network. 
As shown in Fig.\ref{fig:1nn}, this neural network has the structure of 5 inputs representing four normalized state signals and one bias term, a 6-neuron hidden layer, and 2 outputs (one for state-value function and the other for action policy). 

Since only samples of environments are available in the pre-training phase, an off-policy method is employed to train the agent. 
In order to mitigate the bias introduced from the difference between the target policy and the behavior policy, we use the importance sampling, which is defined as:
\begin{equation}\label{eq:rho}
    \rho_t=\frac{\pi(a_t|s_t)}{\pi_b(a_t|s_t)}
\end{equation}
where $\pi(a_t|s_t)$ is the target policy and $\pi_b(a_t|s_t)$ is the behavior policy. 
Together with Eq.(\ref{eq:update_v}) and Eq.(\ref{eq:update_p}), the value parameters $\theta_v$ are updated in direction of
\begin{equation}\label{eq:update_v_off}
    \rho_t\delta_t\nabla_{\theta_v}V_{\pi}(s_t;\theta_v),
\end{equation}
and policy parameters $\theta_p$ are updated in direction of
\begin{equation}\label{eq:update_p_off}
    \rho_t\delta_t\nabla_{\theta_p}\log \pi(a_t|s_t;\theta_p)
\end{equation}
respectively. 
Note that there are some shared parameters in $\theta_v$ and $\theta_p$, and for clarity and simplicity we write them separately.
To improve the sample efficiency and prevent from catastrophic forgetting during training, we also use an experience replay buffer in the pre-training.

\subsubsection{Synchronous Re-training}
We speed up the training process by using a synchronous architecture, as shown in Fig.\ref{fig:parallel}.  
This architecture consists of several actor-learners, each of which interacts with its own environment independently. 
During the re-training, all the actor-learners have the same weights, and at each time step, we compute the gradients of each actor-learner and then these gradients are summed up to perform globally weight updates on all actor-learners.
The detailed synchronous re-training process is shown in Algorithm~\ref{alg:1}. 
With this parallel method, the correlation between consecutive training samples is reduced, and thus a better sample efficiency will be received.

\begin{algorithm}
\SetAlgoLined
\LinesNumbered
Initialize the actor-learner $i$ with pre-trained weights $\theta_v$ and $\theta_p$ \tcp*{Assume $K$ actor-learners are employed in total.}
Initialize time step $t \gets 0$\;
Initialize step counter $count \gets 0$\;
Initialize inverted pendulum state\;
\While{$count<maxstep$}{
  Get state $s_t$\;
  Perform action $a_t$ based on policy $\pi(a_t|s_t;\theta_p)$\;
  Receive reward $r_t$ and new state $s_{t+1}$\;
  \eIf{$r_t\neq-1$}{
   $\delta_t=r_{t}+\gamma {V_\pi}(s_{t+1};\theta_v)-{V_\pi}(s_{t};\theta_v)$\;
   }{
   $\delta_t=r_{t}-{V_\pi}(s_{t};\theta_v)$\;
  }
  Compute gradients wrt $\theta_v$: $d\theta_{v}(i)=\nabla_{\theta_v}V_{\pi}(s_t;\theta_v)\delta_t$\;
  Compute gradients wrt $\theta_p$: $d\theta_{p}(i)=\nabla_{\theta_p}\log \pi(a_t|s_t;\theta_p) \delta_t$\;
  $t \gets t+1$\;
  $count \gets count+1$\;
  \tcc{Next two lines are done in the Gradient summation part shown in Fig.\ref{fig:parallel}(a).}
  Accumulate gradients at $count$ step among all actor-learners wrt $\theta_v$: $d\theta_v={\sum\limits_{i=1}^K d\theta_{v}(i)}$\;
  Accumulate gradients at $count$ step among all actor-learners wrt $\theta_p$: $d\theta_p={\sum\limits_{i=1}^K d\theta_{p}(i)}$\;
  Perform globally synchronous update of $\theta_v$ using $d\theta_v$ and of $\theta_p$ using $d\theta_p$ \tcp*{All actor-learners are updated simultaneously.}
  \If{$r_t=-1$ or $t=5000$}{
   $t \gets 0$\;
   Get a new initial state of pendulum for $s_t$\;
   }
 }
 \caption{Synchronous actor-critic pseudo-code for actor-learner $i$}
 \label{alg:1}
\end{algorithm}

\subsubsection{Training Approach}
Three training approaches are used: (i)~\textit{Baseline} (labeled with 'zero' in following figures), (ii)~\textit{Exact} (labeled with 'pre' in following figures) and (iii)~\textit{Variable Amplitude}. The first two training approaches are the same as described in Section III.A. While in \textit{Variable Amplitude}, pre-training is used and weights can be updated proportionally to gradients. For all cases, $\gamma=0.9$; the learning rate of $\theta_v$ in the output layer and the hidden layer is 0.25 and 0.2, respectively; and the learning rate of $\theta_p$ in the output layer and the hidden layer is 0.2 and 0.1, respectively.

\subsubsection{Simulation Configuration}
In the pre-training, we use an experience replay memory to store 150,000 pre-obtained samples of the environment. All the samples are stored in tuples $(s_t, a_t, r_t, s_{t+1})$, where $s_t$ and $a_t$ are randomly chosen in the state space and the action space, respectively. 
In each training step, one sample is randomly picked from the experience replay memory; and the pre-training is stopped once 200,000 samples are consumed.

\begin{figure}[tbp]
\centering
    \includegraphics[width=3.3in]{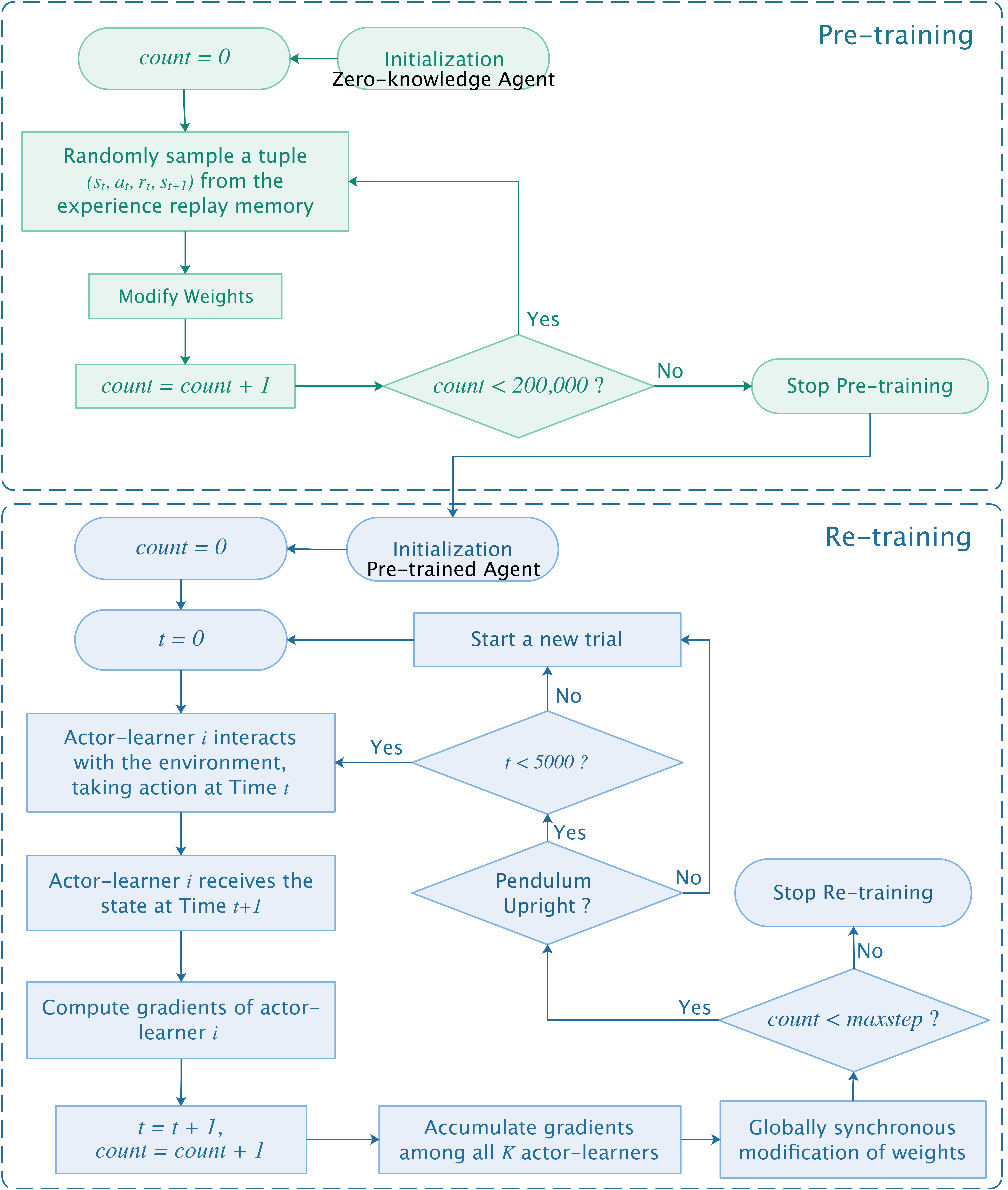}
    \caption{Simulation flowchart when the environmental information is limited in pre-training.}
    \label{fig:limited}
\end{figure}

In the re-training, situations with different number (1, 2, 4 and 8) of actor-learners are considered. To balance parallelism and hardware resource, the 4-actor-learner architecture is employed for the further hardware simulation. 
During the re-training process, the pendulum is manually forced to fall down after kept upright for 5000 time steps (as line 21-24 in Algorithm~\ref{alg:1}), to balance exploitation and exploration. 
The re-training is stopped once 500,000 samples are consumed in total (i.e. $K \times maxstep =500,000$ in Algorithm~\ref{alg:1}), and several checkpoints are inserted to observe the progress during the re-training.
The simulation flow is summarized in Fig.~\ref{fig:limited}.

Besides the total number of updates per weight and $t2f$, we also put an eye on the sample efficiency, which benefits from parallelism in the re-training.

\section{Hardware Implementation}

We consider a mixed-signal circuit system, with crossbar integrated memristors, CMOS neurons and other peripherals for additional functionality (Fig.~\ref{fig:schematic} and Fig.~\ref{fig:parallel}).

\subsection{Separate Action and Evaluation Networks}
\begin{figure}[tbp]
    \centering
    \includegraphics[width=3.5in]{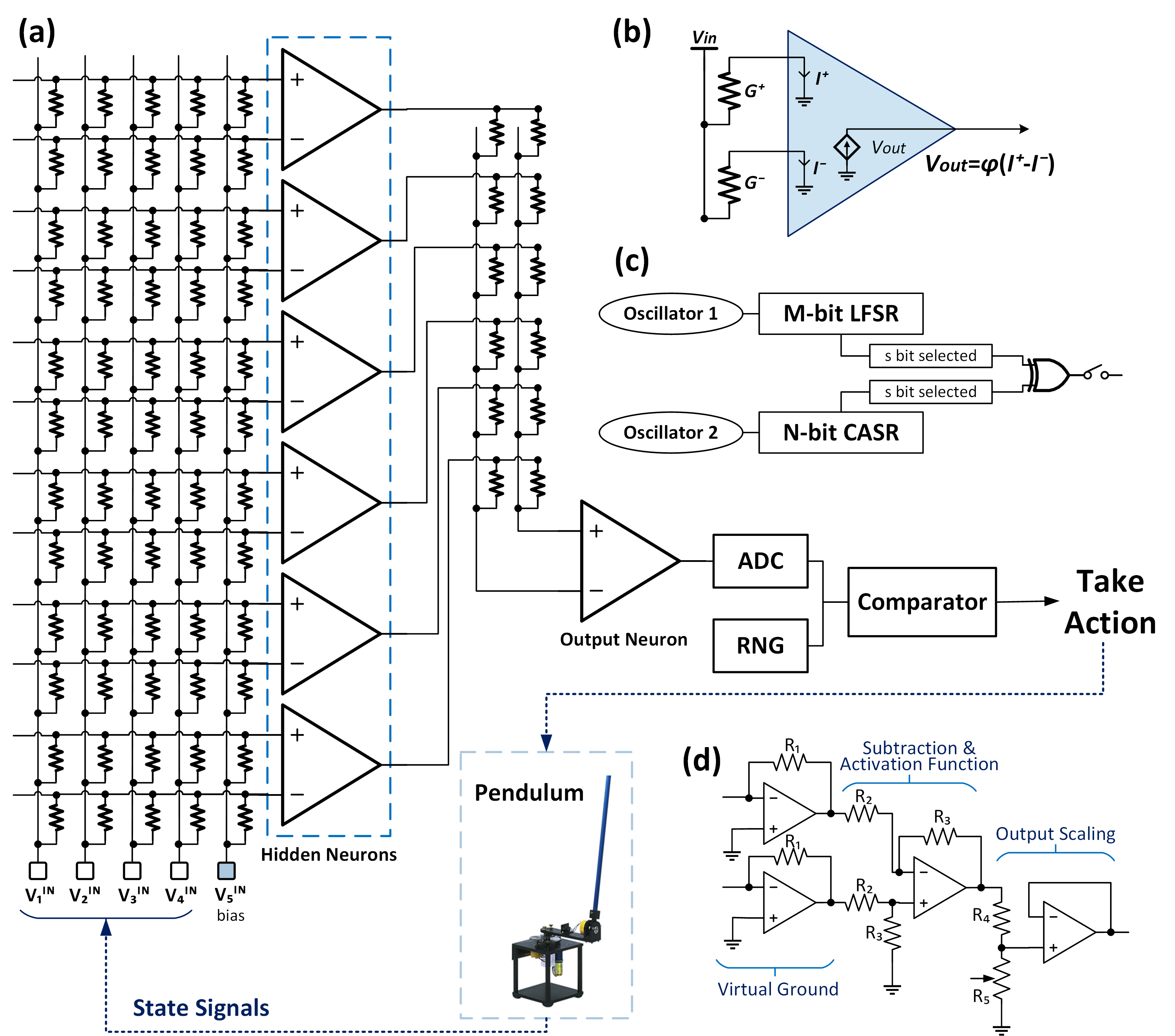}
    \caption{
        System and circuit level implementation details.
        (a) Schematic circuit diagram of the action network, showing its interaction with the pendulum. Note that for the evaluation network, the output neuron does not perform activation function.
        (b) The equivalent circuit showing one differential weight and a neuron with activation function  $\varphi$.
        (c) Block diagram of the hardware random number generator.
        (d) Circuit diagram of the CMOS neuron.
    }
    \label{fig:schematic}
\end{figure}

In the complete environmental information scenario (detailed in Section III.A), two neural networks are employed, i.e., the action and evaluation networks. 
In both networks, state signals are encoded by voltages that are applied to the crossbar circuits.
In the action network, the output voltage of the output neuron represents the probability $p$ to take a CCW push. This voltage is converted into a digital signal and compared to the output of the random number generator to generate the actual action. With the agent taking an action, the pendulum is transitioned into a new state. 
In this new state, the evaluation network, whose circuitry is similar to that of the action network with just few modifications (Fig.~\ref{fig:schematic}(a)), outputs the new state-value $\hat{V_\pi}$.
The temporal difference and weight modifications in two networks are then computed based on the new and previous state values. Finally, a new time step starts by applying new state signals into action and evaluation networks with updated weights.

CMOS neuron realizes the activation function by converting its input current into a voltage (Fig.~\ref{fig:schematic}(d)).
The random number generator (RNG)~\cite{tkacik2002hardware} is implemented by computing exclusive-OR between the outputs of a linear feedback shift register (LFSR) and a cellular automata shift register (CASR) (Fig.~\ref{fig:schematic}(c)). Our simulation results show that 8-bit precision for both the analog to digital converter (ADC) and RNG is sufficient for reaching comparable performance to that of the system with full-precision components.

To achieve better functional performance, the weights are mapped to differential pairs of conductances symmetric to the middle of memristors' conductance range~\cite{prezioso2015training}. For the hardware implementation in this separate-network scenario, we consider \textit{in-situ} training with the Manhattan update rule, which involves applying fixed-magnitude voltage pulses across memristors in four steps~\cite{zamanidoost2015manhattan}. Note that the change in memristor conductance depends on the current state of the device and memristor's effective switching thresholds, when considering device variations, so that the Manhattan update rule is not followed exactly in the hardware implementation. 

\subsection{Actor and Critic Sharing One Network}
\begin{figure}[tbp]

    \centering
    \includegraphics[width=3.4in]{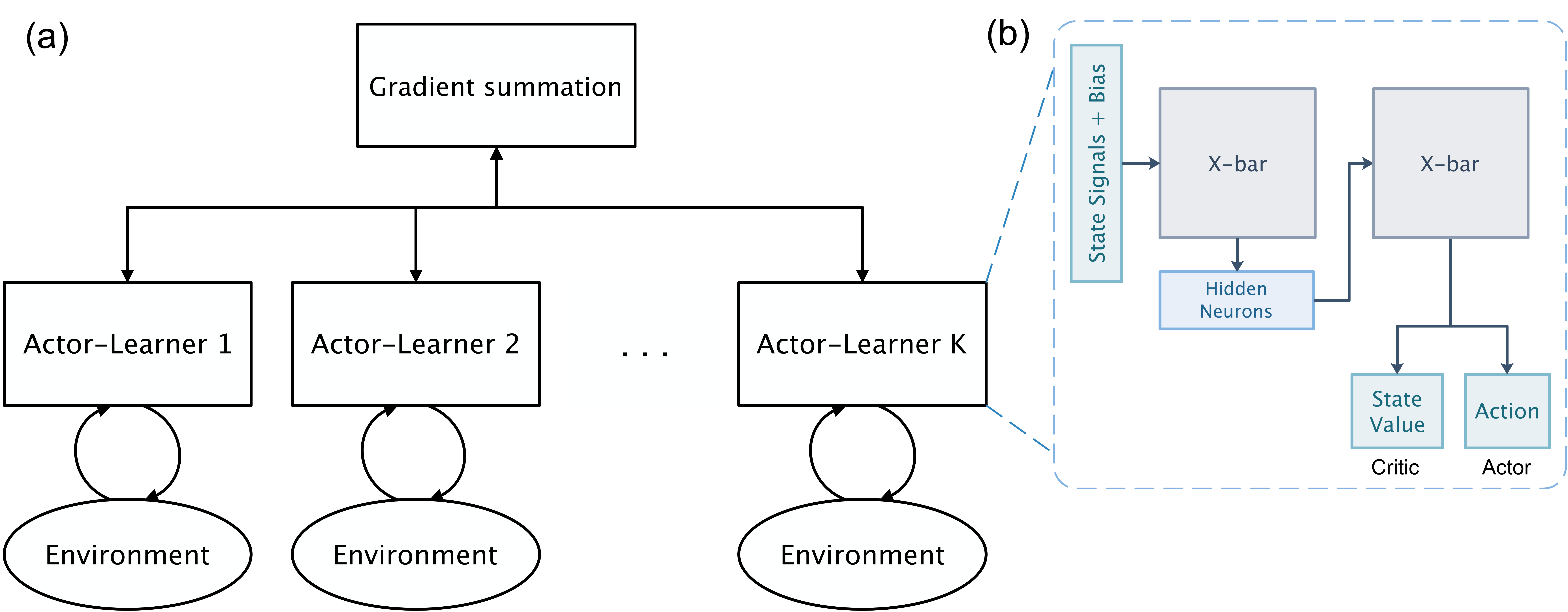}
    \caption{
        System and circuit level implementation for parallel re-training.
        (a) Synchronous training with multiple actor-learners.
        (b) Schematic diagram of one actor-learner
    }
    \label{fig:parallel}
\end{figure}

In the limited environmental information scenario (detailed in Section III.B), a synchronous architecture is used as shown in Fig.\ref{fig:parallel}(a), and each of the actor-learner has the same circuitry (see Fig.\ref{fig:parallel}(b)). 
Similarly, for each actor-learner, state signals are encoded into voltages and applied to the crossbar circuits. 
The output of the actor represents the probability to take a CCW push, and note that CMOS peripherals (ADC, RNG and comparator) are omitted in Fig.\ref{fig:parallel}(b) for brevity. 
By taking the action generated by the actor, the pendulum transitions into a new state and the temporal difference can be derived.
Then weights are updated in a synchronous way (line 18-20 in Algorithm~\ref{alg:1}).

For \textit{in-situ} hardware training in this one-network scenario, we use the \textit{Variable Amplitude} update rule. 
The key idea is to change device conductance proportionally to the weight updates described as
\begin{equation}
    \Delta \textbf{w}=\eta(d\theta_v+d\theta_p),
\end{equation}
where $\eta$ represents the learning rate and denotations are adopted from Algorithm~\ref{alg:1}. 
This can be achieved by applying logarithmically scaled voltages to the crossbar lines~\cite{kataeva2015efficient}. 
Let $V_X$ and $V_Y$ represent the voltages applied to the horizontal and vertical crossbar lines respectively and $\Delta w$ be one element in $\Delta \textbf{w}$. 
In the case $\Delta w>0$, applying
\begin{equation}
    \begin{split}
        & V_X\propto \log [k]\\
        & V_Y\propto -\log [\Delta w/k]
    \end{split}
\end{equation}
results in applying
\begin{equation}
    |V_X|+|V_Y|\propto \log[k\times\Delta w/k]=\log [\Delta w]
\end{equation}
across a selected memristor device. 
Since both \textit{set} and \textit{reset} switching rates are approximately exponential with respect to the applied voltage across the device~\cite{bayat2015phenomenological}, such a method results in the update
\begin{equation}
    \Delta G \propto e^{|V_X|+|V_Y|} \propto \Delta w.
\end{equation}
The pulse duration can be used to control the learning rate and the signs of $V_X$ and $V_Y$ should be flipped for other cases of $\Delta w$. 
By employing the \textit{Variable Amplitude} update rule, memristors in the crossbar are updated line-by-line, since in the proposed synchronous parallel architecture weights are modified according to the summation of gradients of every actor-learner. 
This scheme might cause potentially slower training time as the system scales up.

\section{Result and Discussion}

\subsection{Complete Environmental Information in Pre-training}
\begin{figure}[]
\centering
    \includegraphics[width=3.25in]{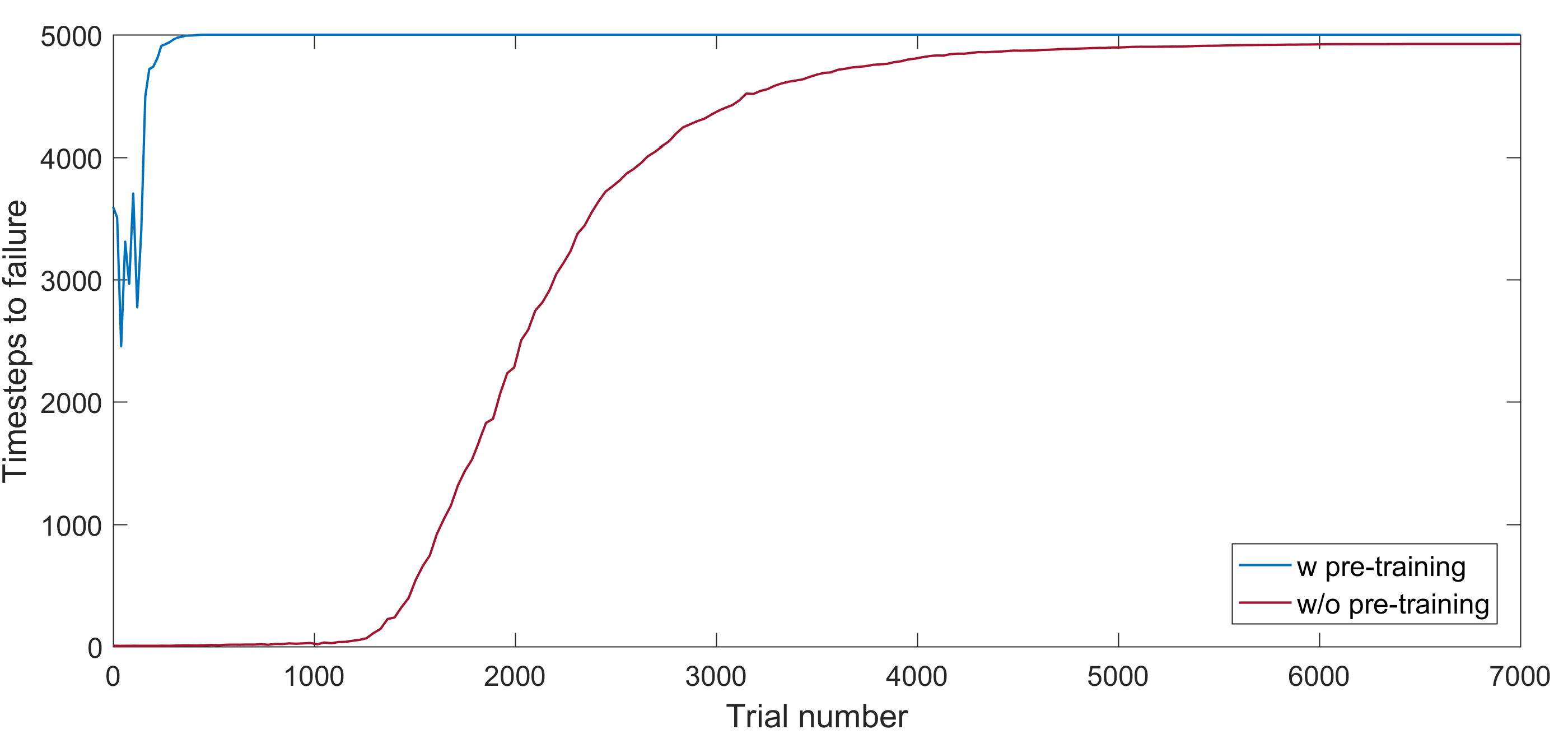}
    \caption{Comparison of learning curves with and without pre-training for the \emph{Exact} training approach. The data are smoothed by averaging time steps per trial into bins of 35 trials.}
    \label{fig:curves}
\end{figure}

\begin{figure}[tp]
    \centering
    \includegraphics[width=3.4in]{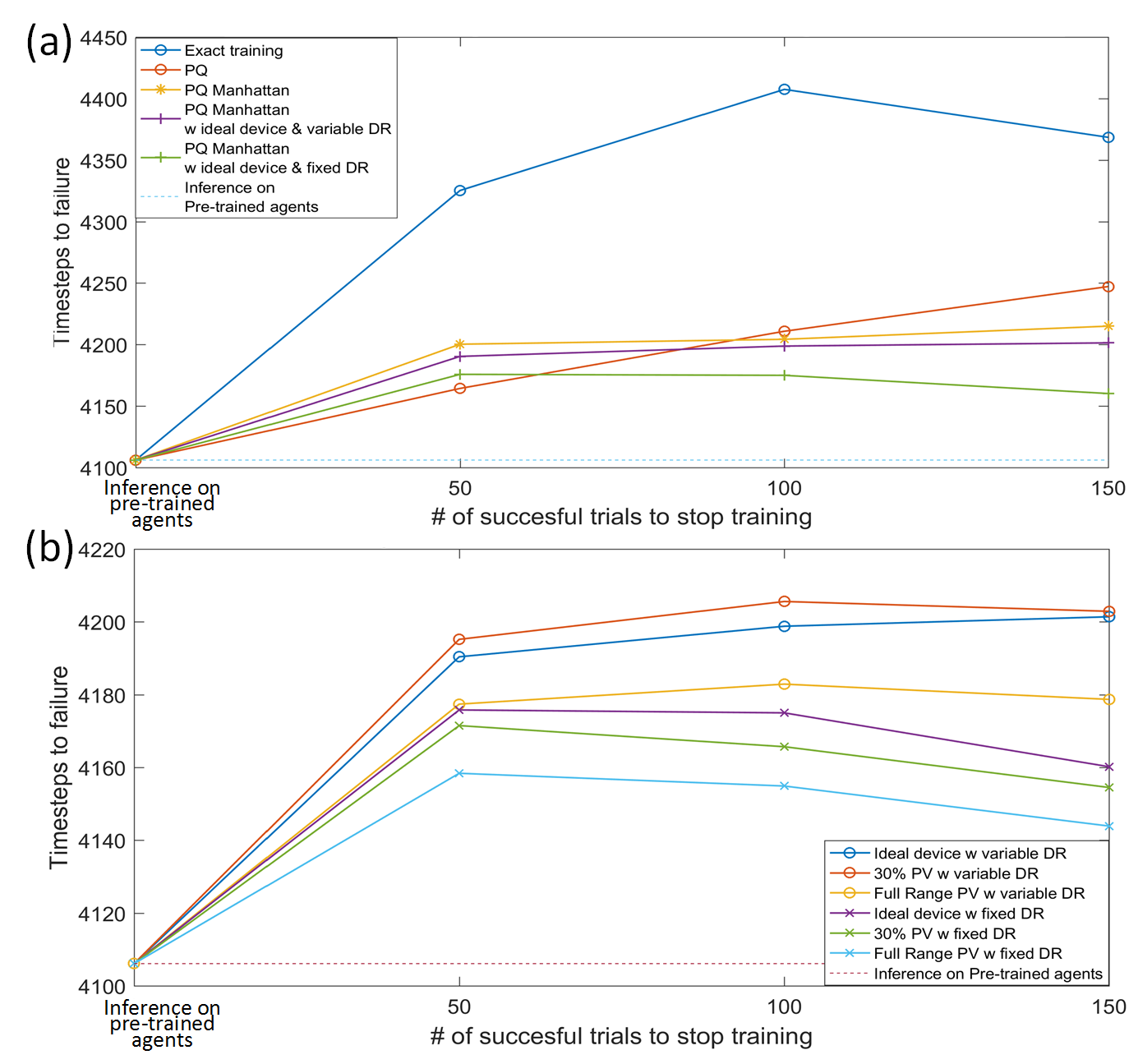}
    \caption{Comparison of time steps to failure for considered training approaches. X-axis represents the number of cumulative successful trials after which the training is stopped. Note that a longer training does not always mean better t2f performance. }
    \label{fig:result1}
\end{figure}

Fig.~\ref{fig:curves} and Table I support the main idea of our approach -  a reduction of training time, and hence the number of weight updates, when using pre-trained weights. 
Agents trained using \emph{Exact} approaches achieve better \textit{t2f} compared to the \emph{Baseline}, however, the number of updates per weight is still prohibitively large for memristor-based hardware implementation. 
PQ training rule can significantly reduce the number of updates per weight during re-training, e.g., reaching comparable to the best case \textit{t2f} performance with only \SI{6.3}{\percent} of the weight updates, cf. \emph{Exact} PQ with $C=400$ and \emph{Exact} with $C=100$. 
PQ modification helps even for the \emph{Baseline}. \emph{Baseline} PQ with $C=50$ achieves the comparable performance to \emph{Baseline} with $C=50$, with 3.13$\times$ fewer weight updates. Requiring less weight updates, \emph{Manhattan} PQ achieves comparable \textit{t2f} for smaller C to that of \emph{Exact} PQ training rule, but its \textit{t2f} saturates early (see Fig.~\ref{fig:result1}(a)).

Table~\ref{tab:efficiency} shows the efficiency of weight updates for the \emph{Manhattan} PQ training. Adjusting discount rate during the training process increases the efficiency of weight updates by as much as \SI{102}{\percent}. \textit{t2f} is only slightly reduced when assuming significant variations in the device behaviour (Fig.~\ref{fig:result1}(b)). In fact, a slight amount of process variation may result in a better performance, since adding proper noise in parameter space can benefit policy exploration \cite{plappert2017parameter}.

\begin{table}[htbp]
\centering
\caption{\# Updates per Weight and Time steps to Failure, averaged on 2500 agents.}
\begin{tabular}{c|c|c|c}
\hline
Training Approach & \begin{tabular}[c]{@{}c@{}} C \end{tabular} & \begin{tabular}[c]{@{}c@{}}\# Updates per Weight\end{tabular} & \begin{tabular}[c]{@{}c@{}} t2f \end{tabular} \\ \hline
Baseline & 50                                                                 & 3.5610e+05              & 4159.1                                                                   \\
Baseline PQ & 50                                                              & 1.1360e+05              & 4062.6                                                                   \\
\rowcolor[HTML]{BDCBDF} 
\begin{tabular}[c]{@{}c@{}}Inference on \\pre-trained agent\end{tabular} & None & None                    & 4106.1                                                                   \\
\rowcolor[HTML]{DAE8FC} 
Exact & 50                                                                    & 2.6249e+05              & 4325.3                                                                   \\
\rowcolor[HTML]{DAE8FC} 
Exact & 100                                                                   & 5.1997e+05              & 4407.5                                                                   \\
\rowcolor[HTML]{DAE8FC} 
Exact PQ & 100                                                                      & 8.9290e+03              & 4210.9                                                                   \\
\rowcolor[HTML]{DAE8FC} 
\rowcolor[HTML]{DAE8FC} 
Exact PQ & 400                                                                      & 3.2968e+04              & 4338.9                                                                   \\
\rowcolor[HTML]{DAE8FC} 
Manhattan PQ (ideal) & 50                                                             & 1.8256e+03              & 4200.3                                                                   \\
\rowcolor[HTML]{DAE8FC} 
Manhattan PQ (ideal) & 100                                                            & 3.5108e+03              & 4204.5                                                                   \\ \hline
\end{tabular}
\label{tab:performance}
\end{table}

\begin{table}[htbp]
\caption{Weight Update Efficiency for Manhattan Approach.}
\vspace{-0.1in}
\centering
\begin{tabular}{c|c|c|c|c}
\hline
\rowcolor[HTML]{BCDDEC} 
Training Approach             & Device Assumptions & C=50   & \multicolumn{1}{l|}{\cellcolor[HTML]{BCDDEC}C=100} & \multicolumn{1}{l}{\cellcolor[HTML]{BCDDEC}C=150} \\ \hline
                              & Ideal devices      & 0.0518 & 0.0294                                             & 0.0205                                            \\
                              & 30\% PV            & 0.0549 & 0.0316                                             & 0.0208                                            \\
\multirow{-3}{*}{Variable DR} & Full range PV      & 0.0433 & 0.0242                                             & 0.0155                                            \\ \hline
                              & Ideal devices      & 0.0424 & 0.0216                                             & 0.0115                                            \\
                              & 30\% PV            & 0.0397 & 0.0187                                             & 0.0103                                            \\
\multirow{-3}{*}{Fixed DR}    & Full range PV      & 0.0319 & 0.0153                                             & 0.0080                                            \\ \hline
\end{tabular}
\label{tab:efficiency}
\end{table}

\begin{figure*}[bh]
\centering
    \includegraphics[width=6.8in]{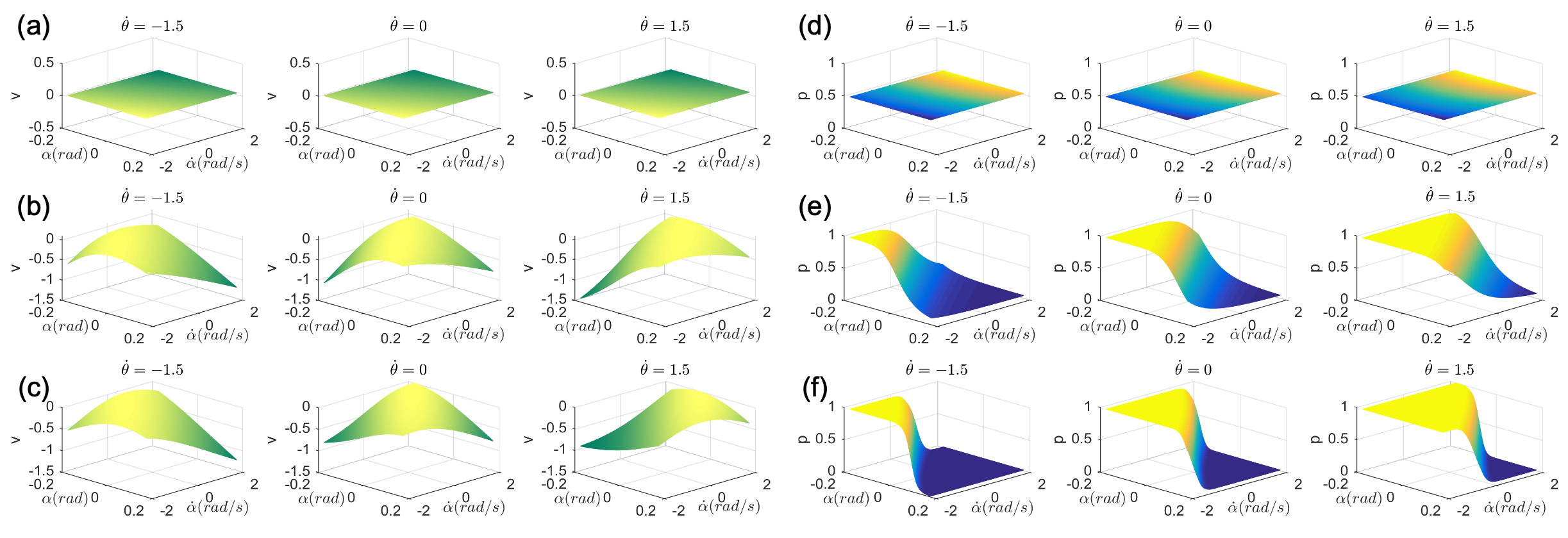}
    \caption{The value function ((a), (b) and (c)) and the policy ((d), (e) and (f)) as projections onto $\alpha$ and $\dot{\alpha}$ plane. As the rotary inverted pendulum is symmetric on the rotary arm angle $\theta$, $\theta=0$ is considered with three different $\dot{\theta}$ values in this figure. Zero-knowledge agents: (a), (d); after pre-training: (b) and (e); after re-training: (c) and (f). }
    \label{fig:projection}
\end{figure*}

\subsection{Limited Environmental Information in Pre-training}

Fig.\ref{fig:projection} shows the graphical representation of the outputs of the actor and critic neurons (i.e., the policy and the state-value function), with weights at different learning stages (i.e., zero-knowledge, after pre-training, and after re-training), which are projections onto the $\alpha$, $\dot{\alpha}$ plane with three different values of $\dot{\theta}$, respectively.
As the learning progresses, the policy and the state-value function evolve.
As depicted in Fig.\ref{fig:projection}(b)(e), the outputs after pre-training already have shapes similar to those of the learned policy and state-value function, which indicates that agents do learn some basic knowledge from pre-traing, even with limited environmental information.
In Fig.\ref{fig:projection}(c), the learned value function has a helpful and intuitive output that forms a diagonal ridge with low values at $+\alpha$, $+\dot{\alpha}$ and $-\alpha$, $-\dot{\alpha}$, i.e., the region of the state space that is likely to fall. 
For most of the times, as illustrated in Fig.\ref{fig:projection}(f), the actor generates a probability very close to either 0 or 1, indicating CW pushes or CCW pushes, with a relatively narrow transition zone that the action selection is less deterministic. These transition zones correspond to the diagonal ridges shown in Fig.\ref{fig:projection}(c). 

\begin{figure*}[!htbp]
\centering
    \includegraphics[width=6.8in]{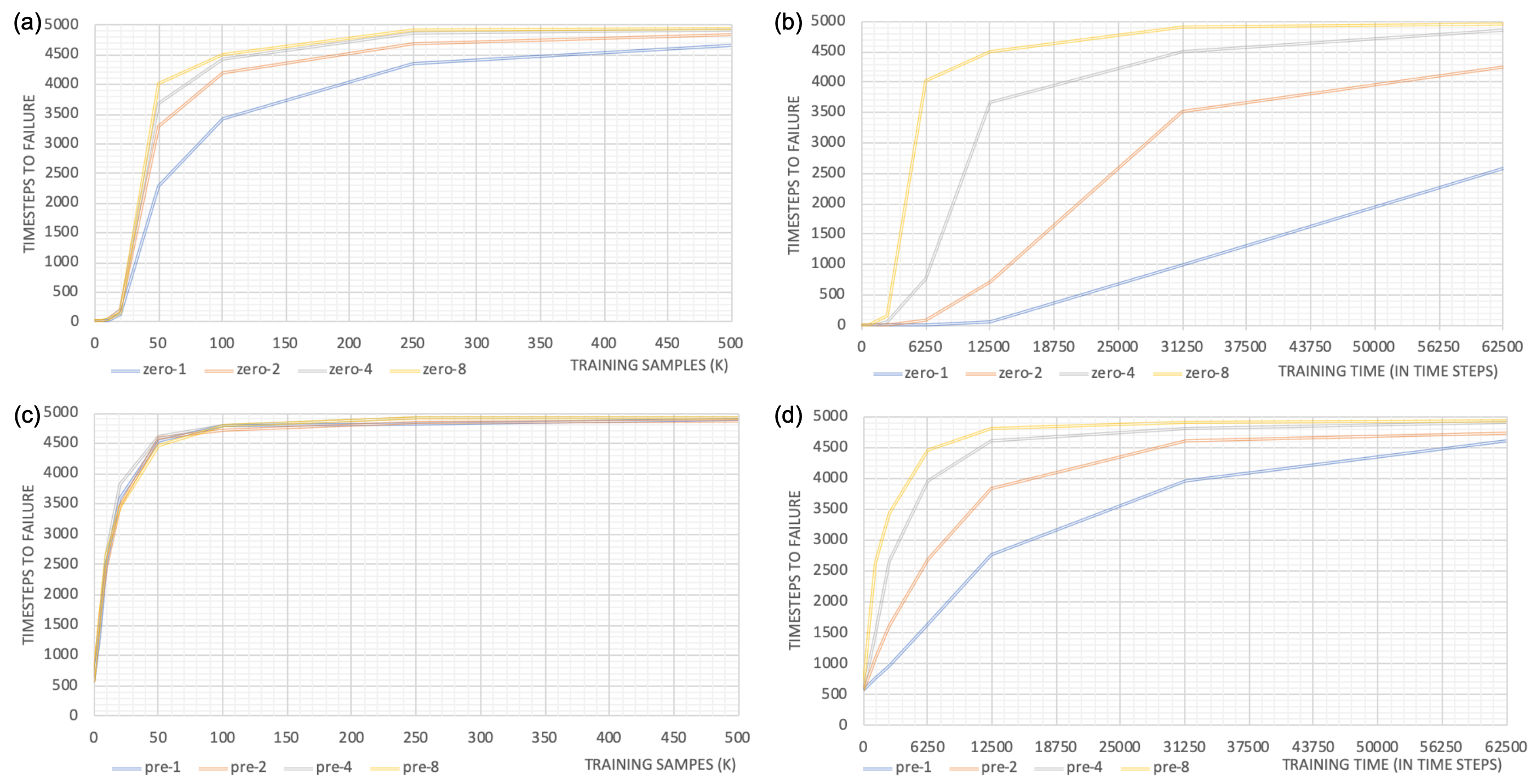}
    \caption{Sample efficiency and learning speed comparisons of different numbers of actor-learners with and without pre-training. In (a) and (c), the x-axis shows the total number of training samples in thousands; in (b) and (d), the x-axis shows the training time in time steps; and in all cases, the y-axis shows the averaged t2f of 100 agents on the test set. Since weights are updated at every time step, the x-axis in (b) and (d) can also be recognized as the number of updates per weight. Denotations in the figure: 'zero-$i$' and 'pre-$i$' for zero-knowledge agents and pre-trained agents with $i$ parallel actor-learners, respectively.}
    \label{fig:result2}
\end{figure*}

Fig.\ref{fig:result2}(b) and (d) compare the learning speed among different numbers of parallel actor-learners. 
Obviously, we can achieve better speedups with greater numbers of parallel actor-learners. 
The same statement can be reached that with pre-trained weights, there is a significant reduction of training time. The gap of learning speed between zero-knowledge agents and pre-trained agents is shrinking as the number of parallel actor-learners is increasing, which conveys an insight that if there is abundant hardware resource, better parallelism can partly take the role of pre-training. Fig.\ref{fig:result2}(a) and (c) show the sample efficiency comparison of different numbers of actor-learners. Unchanged or better sample efficiency is obtained from more actor-learners, which guarantees linear or super-linear speedups with increased number of actor-learners.

\begin{figure}[!htbp]
\centering
    \includegraphics[width=3.4in]{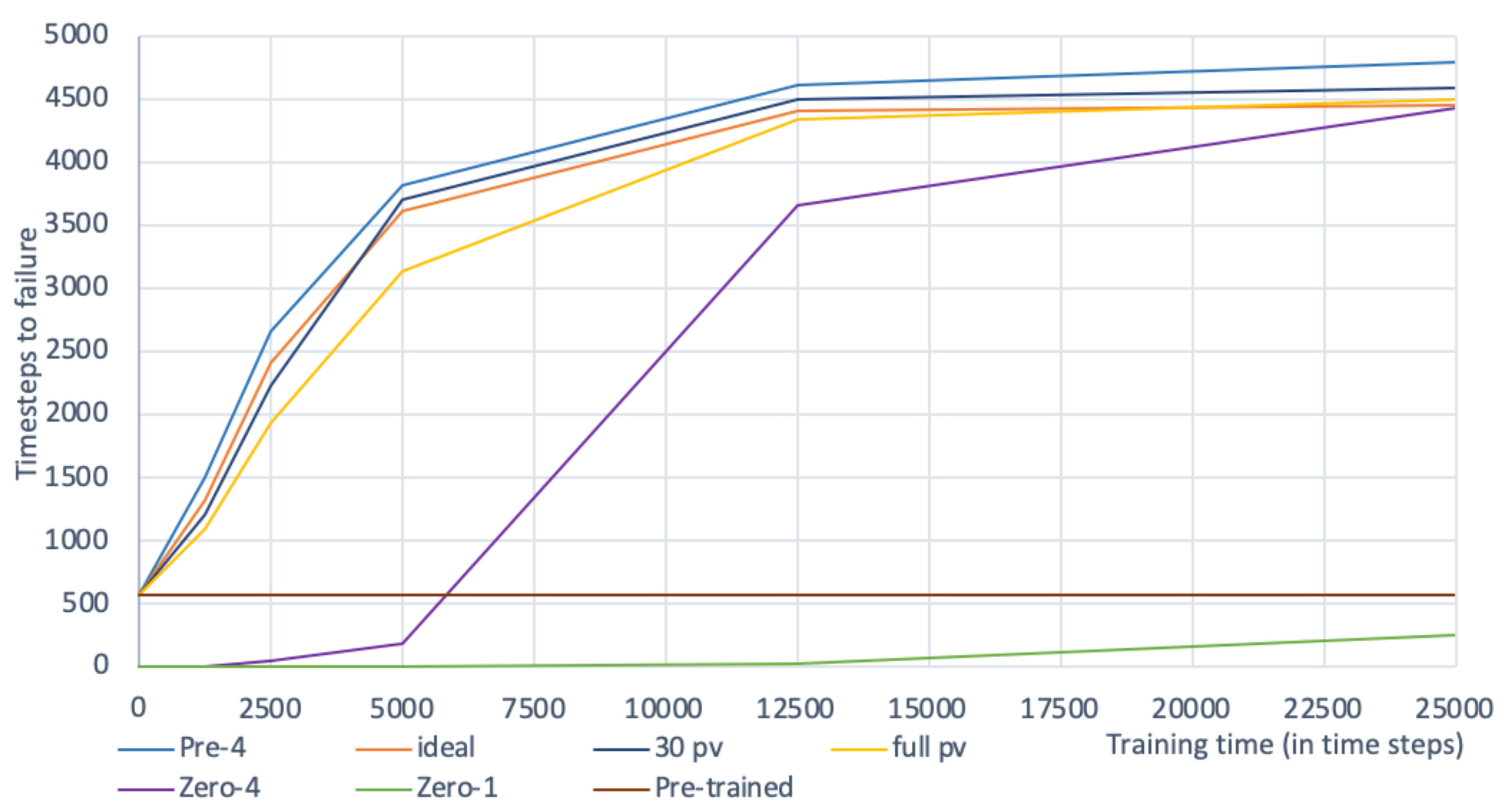}
    \caption{Comparisons of time steps to failure for different scales of process variation.}
    \label{fig:hwresult}
\end{figure}

Simulation results with device models are illustrated in Fig.\ref{fig:hwresult}. When a large process variation is assumed, there is a slight degradation of $t2f$ in the early learning stage, but as training continues it can be mitigated. Generally, around 3$\times$ $\sim$ 4$\times$ and 30$\times$ $\sim$ 40$\times$ fewer updates per weight are required to reach comparable $t2f$ performance in comparison to zero-4 and zero-1, respectively. This indicates the importance to employ pre-training and parallel re-training.

\section{Conclusion}
In this work we study a neuromorphic hardware implementation for reinforcement learning based on memristive circuits, focusing on the well-known inverted pendulum benchmark. 
Due to a large number of weights updates, which are required to converge in the reinforcement learning algorithm, training from zero-knowledge agents is currently impractical for memristor-based hardware.
Based on this consideration, we propose a two-fold training procedure: \textbf{\textit{ex-situ} pre-training and \textit{in-situ} re-training}.
With this training procedure, two scenarios are considered based on the difficulty to obtain environmental information:
\begin{itemize}
    \item If environmental information is easy to obtain in pre-training (i.e., complete environmental information), the re-training can start with a rather good pre-training, and we suggest several more practical training techniques. The most important result is that \emph{Manhattan} PQ rule with pre-training, a practical approach for in-situ learning hardware implementation, achieves comparable functional performance to that of \emph{Baseline} method, while requiring 195x fewer weight updates.
    Furthermore, to deal with memristors' non-idealities, we propose to adjust the discount rate in temporal difference learning algorithm during the training and show that scheme increases efficiency of weight updates by more than \SI{100}{\percent}.
    \item If environmental information is limited in pre-training, we propose an off-policy method with experience replay and importance sampling in pre-training, to improve sample efficiency and reduce the bias; and in re-training we recommend to use a synchronous parallel architecture to further reduce the updates per weight in the training process. In this case, an advanced approach for in-situ learning hardware implementation, \textit{Variable Amplitude}, is applied. 
    Results show that with these proposed methods, 30$\times$ $\sim$ 40$\times$ fewer weight updates is required to achieve comparable performance to that of zero-1.
\end{itemize}

We believe that our results show promise for implementing fast and energy-efficient memristor-based neuromorphic hardware for reinforcement learning to address complex cognitive tasks.


\bibliographystyle{IEEEtran}
\bibliography{ref}

\end{document}